\pdfoutput=1

\documentclass[11pt,twoside,a4paper,cmspaper,final,collab]{cms-tdr}
\def\svnVersion{168105}\def\cmsCernNo{2012-278}\def\cmsCernDate{\today}\def\cmsMessage{Submitted to Physics Letters B}

\begin{document}\cmsNoteHeader{EXO-11-073}

\hyphenation{had-ron-i-za-tion}
\hyphenation{cal-or-i-me-ter}
\hyphenation{de-vices}

\RCS$Revision: 165936 $
\RCS$HeadURL: svn+ssh://svn.cern.ch/reps/tdr2/papers/EXO-11-073/trunk/EXO-11-073.tex $
\RCS$Id: EXO-11-073.tex 165936 2013-01-21 15:05:40Z alverson $
\def\tt{\ttbar}
\newcommand{\MSig}{\ensuremath{M_\Sigma}\xspace}
\newlength\cmsFigWidth
\ifthenelse{\boolean{cms@external}}{\setlength\cmsFigWidth{0.85\columnwidth}}{\setlength\cmsFigWidth{0.6\textwidth}}

\cmsNoteHeader{CMS-11-073} 
\title{Search for heavy lepton partners of neutrinos in proton-proton collisions in the context of the type III seesaw mechanism}

\date{\today}

\abstract{
A search is presented in  proton-proton collisions  at $\sqrt{s}=7$\TeV
for fermionic triplet states expected in type III seesaw models.
The search is performed using final states with three isolated charged leptons and an imbalance in transverse momentum.
The data, collected with the CMS detector at the LHC, correspond to  an
integrated luminosity of 4.9\fbinv.
No excess of events is observed above the  background predicted
by the standard model,
and the results are interpreted in terms of limits
on production cross sections and masses of the heavy partners of the neutrinos in type III seesaw models.
Depending on the considered scenarios, lower limits are obtained on the mass of the heavy partner of the neutrino that range from 180 to 210\GeV.
These are the  first limits on the production of type III seesaw fermionic
triplet  states reported by an experiment at the LHC.
}

\hypersetup{%
pdfauthor={CMS Collaboration},%
pdftitle={Search for heavy lepton partners of neutrinos in proton-proton collisions in the context of the type III seesaw mechanism},%
pdfsubject={CMS},%
pdfkeywords={CMS, physics}}

\maketitle 

\section{Introduction \label{s:intro}}

Experiments on  neutrino oscillations~\cite{2010er,2011zk,2012eh,2012reno}
indicate that neutrinos have mass
and their masses are much smaller than those of the charged leptons.
However, the origin of neutrino mass is still unknown.
An interesting possibility
 is provided by the seesaw mechanism,
in which a small Majorana  mass can be generated for each of the known neutrinos
by introducing massive
states  with  Yukawa
couplings to leptons and to the Higgs field.
Seesaw models  called type I~\cite{typeI_,1979ia},
type II~\cite{typeII_,1980gr,1981bx,1980nt,1980qt}, and
type III~\cite{typeIII_,typeIIIma} introduce
heavy states of mass $M$, that involve, respectively,
weak-isospin singlets, scalar
triplets, and fermion triplets.
The neutrino masses are generically reduced relative
to charged fermion masses
by a factor  $v/M$,
where
$v$ is the vacuum expectation value of the Higgs field.
For sufficiently large
$M$ (of the order of $10^{14}$\GeV), small neutrino masses
are generated even for Yukawa couplings of ${\approx}1$.
On the other hand,
either smaller Yukawa
couplings or
extended seesaw mechanisms, such as those of the inverse seesaw models~\cite{delaguila2}, are required
to obtain small neutrino masses while keeping
$M$ close to a few hundreds of \GeV.
At the Large Hadron Collider (LHC), type II and III
states can be produced through gauge
interactions, so that the possible smallness of the Yukawa couplings does not affect
the production cross section of the heavy states.
In particular, the possibility of discovering a type III fermion at  a proton-proton
centre-of-mass  energy of $\sqrt{s}=14$\TeV is discussed in
Refs.~\cite{delaguila_,franceschini_,bajc}. Recently, a leading-order (LO) computation
of the signal expected at  $\sqrt{s}=7$\TeV has become available as a computer program
for simulating such final states~\cite{carflo_}.

\begin{figure}[htb]
    \begin{center}
\includegraphics[width=\cmsFigWidth]{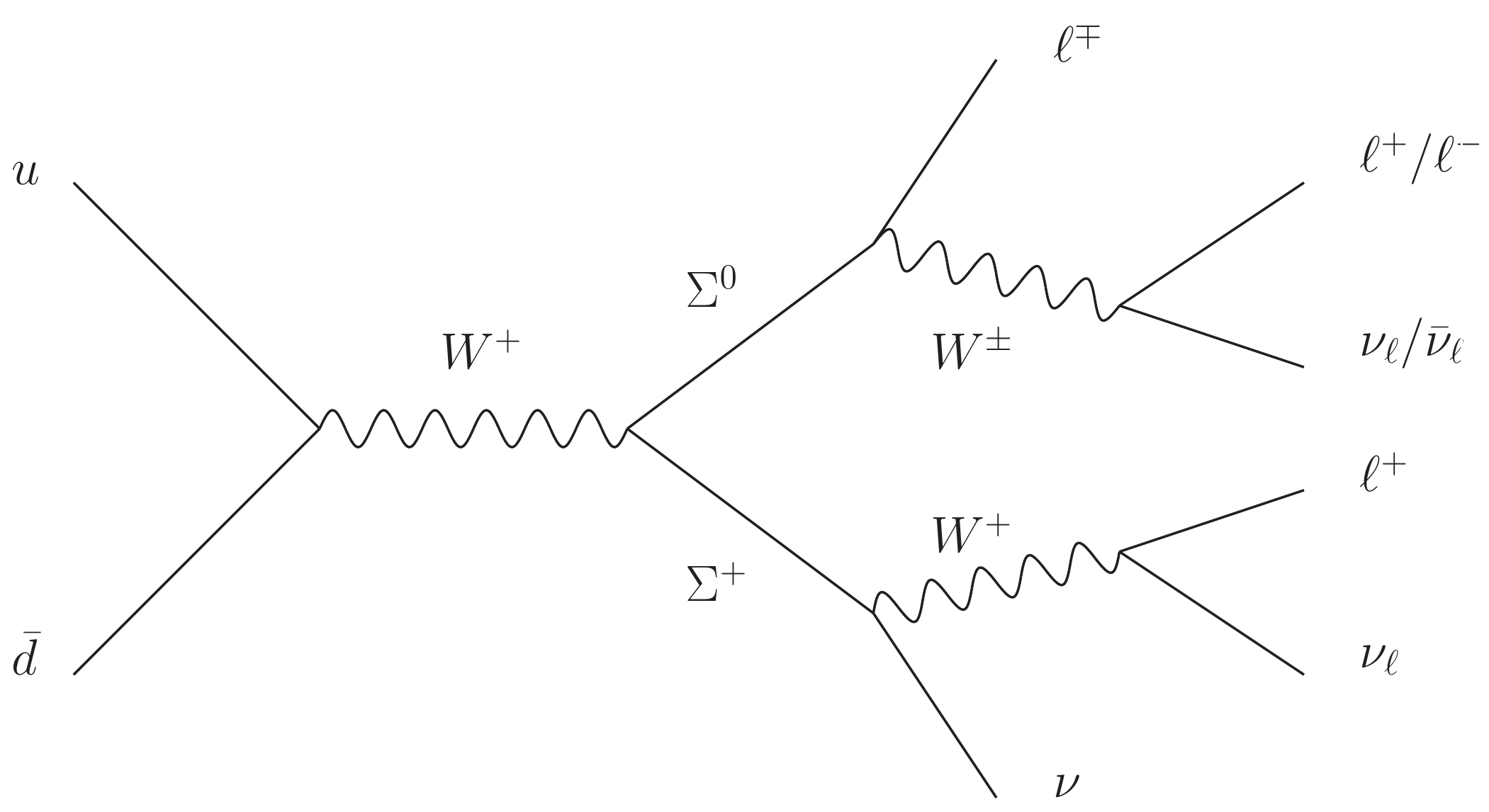}
\caption{Feynman diagram for the dominant contribution to three-charged-leptons final states in pair production of
$\Sigma$ in the type III seesaw models.
The production cross section for the charged-conjugate intermediary  $\PW^-$ is expected to be about a factor of two smaller.
}
        \label{f:Fey}
    \end{center}
\end{figure}

Given the
electric charges  of the lepton triplet,
hereafter referred to as $\Sigma^+$, $\Sigma^0$, and $\Sigma^-$, the most promising
signature  for finding a $\Sigma$ state with a mass $\MSig$
of the order of a few hundreds of\GeV is in production through
quark-antiquark annihilation
$\cPq \cPaq^\prime \to \Sigma^0 \Sigma^+$, followed by the decays
$ \Sigma^0 \to \ell^{\mp}\PW^{\pm} $
and  $ \Sigma^+ \to \PWp \nu$.
The mass differences among the three electric charge states are assumed to be negligible.
The mass range relevant for this analysis is bounded by the present lower limits
(${\approx}100$\GeV) from the L3 experiment~\cite{L3} and by the CMS loss of sensitivity near ${\approx}200$\GeV
because of the very steep decrease of the expected cross section
with mass.
Since there are twice as many u as d valence quarks
in the proton, the production of $\Sigma^+$ $\Sigma^0$ via virtual $\PW^+$ bosons in the $s$-channel (Fig. \ref{f:Fey})
has the highest cross section of all the $\Sigma$ charge combinations.
(The cross section for the charge conjugate intermediary  $\PW^-$ is expected to be about a factor two smaller.)
Selecting  $\PW^{\pm}\to \ell^{\pm} \nu$  decays
(where $\ell$ is an electron or muon) as the final states for the search, offers a very clean
signature of three charged, isolated leptons.
The decay  $ \Sigma^+ \to  \ell^+$Z, with
Z$\to\cPgn\cPagn$ or Z$\to \cPq \cPaq$,
can also contribute significantly to the three-lepton final state,
especially since its relative yield grows with $\MSig$.
The $\tau$ lepton  also contributes  to the three-lepton final states through  $\tau \to \ell \nu_{\ell} \nu_{\tau}$ decays.
Details of the phenomenology
and the different contributions to the final state of interest can be found in Ref.~\cite{carflo_}.

The total width of the $\Sigma$ states and their decay branching fractions
to SM leptons depend
on the mixing matrix element for the
leptons
$V_{\alpha}$, where $\alpha$ labels
each of the $\Pe$, $\mu$, and $\tau$
generations of leptons.
Constraints on the mixing parameters and their products are
available in Refs.~\cite{2007ux,carflo_}.

The $\Sigma \Sigma$ production cross section
does not depend on the matrix elements $V_\alpha$, which enter only in the $\Sigma$ decays.
The fraction of $\Sigma$ decays to the lepton $\alpha$  is proportional to:

\begin{equation}
\label{eq:br}
b_{\alpha} = \frac{|V_{\alpha}|^2}{|V_{\Pe}|^2+|V_{\mu}|^2+|V_{\tau}|^2}.
\end{equation}

If all three $V_{\alpha}$ values are less than ${\approx}10^{-6}$, the $\Sigma$ states
can have sufficiently long lifetimes to produce leptons
at secondary vertices,
a possibility not considered in this analysis.

This Letter reports on a search for fermionic triplet states expected in type III seesaw models, in
final states with three charged leptons and
an imbalance in transverse momentum ($\MET$).
The data sample corresponds to an integrated luminosity of 4.9\fbinv, collected in proton-proton collisions at $\sqrt{s}=7$\TeV with the Compact Muon Solenoid
(CMS) detector at the LHC in 2011. The analysis is based on the model described in Ref.~\cite{delaguila_}, using the implementation of Ref.~\cite{carflo_}.
Three possibilities are considered for the ratios $b_{\alpha}$, defined in Eq.(\ref{eq:br}): first, $b_{\Pe}=b_{\mu}=b_{\tau}=1/3$,
hereafter referred to as the flavor-democratic scenario (FDS), second, $b_{\Pe} = 0$, $b_{\mu}= 1$,
$b_{\tau}=0$, and third, $b_{\Pe}=1$
and $b_{\mu}=b_{\tau}=0$, hereafter referred to as the muon scenario ($\mu\text{S}$) and the electron scenario (eS), respectively.

\section{The CMS detector \label{s:detsim} }

A detailed description of the CMS detector can be found in Ref.~\cite{cmstdr}.
The central feature of the CMS apparatus is a superconducting
solenoid that provides an axial magnetic field of 3.8\unit{T}.
A silicon tracker, a lead-tungstate crystal electromagnetic
calorimeter (ECAL), and a brass/scintillator hadron calorimeter (HCAL) reside within the magnetic field volume.
Muons are
identified using the central tracker and a muon system consisting of gas-ionization detectors embedded in the
steel return yoke outside of the solenoid.

The directions of particles in
the CMS detector are described using the azimuthal angle $\phi$ and the
pseudorapidity $\eta$, defined as $\eta = -\ln[\tan(\theta/2)]$,
where $\theta$ is the polar angle relative to the anticlockwise proton beam.
All objects are reconstructed using a particle-flow (PF)
algorithm~\cite{PFT-09-001,PFT-10-001,PFT-10-002}. The PF algorithm
combines information from all subdetectors to identify and reconstruct particles
detected in the collision, namely charged hadrons, photons, neutral hadrons, muons, and electrons.
Jets are reconstructed using the anti-\kt jet clustering algorithm  with a distance parameter
of 0.5~\cite{anti-kT}.
Jet energies are
corrected
for non-uniformity in calorimeter
response and for differences found between
jets in simulation and in data~\cite{2011ds}.
An imbalance in transverse momentum ($\MET$)
is defined by the magnitude
of the vectorial sum of the transverse momenta ($\pt$) of all particles reconstructed through the PF algorithm.

\section{ Simulation of signal and background \label{s:sim} }

To estimate signal efficiency, $\Sigma^{+}\Sigma^{0}$ events are generated using the \textsc{FeynRules} and \MADGRAPH computer programs described in
Ref.~\cite{carflo_}, while parton showers and hadronization are implemented using the \PYTHIA generator (v6.420)~\cite{PYTHIA} .
The detector simulation is based on the \GEANTfour program~\cite{GEANT4}.
Given the
number of $\MSig$ mass points to be generated, part of the detector
simulation is performed using the CMS Fast Simulation framework~\cite{fast1,fast2}.
Several background sources are considered in this analysis,
the most relevant one being
$\PW$Z production with both bosons decaying into leptons.
A smaller contribution to the background comes from ZZ production, where the Z bosons decay leptonically,
and one of the leptons is  either outside of the detector acceptance or
is misreconstructed.
These two-boson events, calculated at next-to-LO with MCFM~\cite{2010ff}, are generated
with \PYTHIA.
Backgrounds from the production of three EW bosons are generated with \MADGRAPH5~\cite{Madgraph}.
Backgrounds
from jets and photons that are
 misidentified as
leptons are also taken into account,
including events from Drell--Yan $\ell^{+} \ell^{-}$+jets sources ~\cite{PDG_},
 $\PW$+jets, Z+jets,
$\tt$, and Drell--Yan $\ell^{+} \ell^{-}$+$\gamma$ conversions to $\ell^{+} \ell^{-}$.
(The Drell--Yan process consists of $\cPq\cPaq \to \gamma^*/\rm{Z} \to \ell^{+} \ell^{-}$ production, with $\gamma^*$ and $\cPZ$ intermediaries representing virtual $\gamma$ or Z bosons.)

The presence of additional simultaneous pp interactions (pileup) is incorporated by simulating and mixing additional interactions
with a multiplicity matching that observed in data.
\section{ Event selection criteria \label{s:datasets} }

The online trigger and the offline selection criteria are analogous to those used in other multi-lepton analyses performed by the CMS Collaboration~\cite{sus13,exo11041}.
Events are selected through two-lepton triggers in which two muons, two electrons,
or one electron and one muon are required to be present. Because of the steady increase in
 instantaneous luminosity in 2011, some of the lepton $\pt$ thresholds were increased over time
to keep the trigger rates within the capabilities of the data acquisition system.
For the two-muon trigger, the $\pt$ requirements evolved from 7\GeV for each muon to asymmetric
requirements of 17\GeV for the highest-$\pt$ (leading) muon and 8\GeV for the second-highest $\pt$
muon. For the two-electron trigger, the requirement is asymmetric, with a threshold applied
to the energy of an ECAL cluster projected onto the plane transverse to the
beam line
($\et=E \sin \theta$). The cluster of the leading electron is required to have $\et> 17$\GeV, and that of the next-to-leading
electron to have $\et > 8$\GeV. For the electron-muon trigger, the thresholds are either  $\et> 17$\GeV for
the electron and $\pt>8$\GeV for the muon, or $\et > 8$\GeV for the electron and $\pt > 17$\GeV for
the muon.
The selected events must contain
at least two lepton candidates
with trajectories that have a transverse impact parameter of less than 0.2\unit{mm} relative to the principal interaction vertex.
The chosen vertex is defined as the one with the largest value for the sum of the $\pt^{2}$ of the emanating tracks.

Muon candidates are reconstructed from a fit performed to hits in both
the silicon tracker and the outer muon detectors, thereby defining a "global muon".
The specific selection
requirements for a muon are:
(i) $\pt>$ 10\GeV,
(ii) $|\eta| < 2.4$,
(iii)  more than 10 hits in the silicon tracker, and
(iv) a global-muon fit with  $\chi^2 /\text{dof}<10 $, where $\text{dof}$ is the number of degrees of freedom.

Electron candidates are reconstructed using clusters of energy
depositions in the ECAL that match the extrapolation of a reconstructed track.
The electron track is fitted using a Gaussian-sum filter ~\cite{gsf},
with the algorithm taking into account the emission of bremsstrahlung photons in the silicon tracker.
The  specific requirements for a reconstructed electron are:
(i) $\pt>10\GeV$,
(ii) $|\eta| < 1.44$,
within the fully instrumented part of the central barrel, or $1.57 < |\eta| < 2.5$ for the endcap regions,
(iii) not being a candidate for photon conversion, and
(iv) the tracks reconstructed using three independent algorithms~\cite{PFT-10-001} to give the same sign for the electric charge.

All accepted lepton candidates are required to be isolated from
other particles. In particular,  selected muons must have $(\sum \pt)/\pt^{\mu}<0.15$,
where  the sum over scalar $\pt$ includes all other PF objects
within a cone of radius  $\Delta R = \sqrt{(\Delta\eta)^2 +(\Delta\phi)^2}=0.3$
of the muon track,
where $\Delta\eta$ and $\Delta\phi$ are the differences in
pseudorapidity and azimuthal angle between the lepton axis and
the positions of other particles.
Similarly, an electron
candidate
is accepted if $(\sum \pt)/\pt^{\Pe}<0.20$ within a cone of $\Delta R=0.3$.

The  candidate events used for the search are required to have:
(i) three isolated charged leptons originating from the same primary vertex, as defined above,
(ii) sum of the lepton charges equal to $+1$,
(iii) $\MET>30$\GeV,
(iv) $\pt > 18, 15, 10$\GeV for the lepton of highest, next-to-highest, and lowest $\pt$, and
(v) $H_{\mathrm{T}} < $ 100\GeV,
where $H_{\mathrm{T}}$
is the scalar sum of the transverse momenta  of jets with   $\pt > 30\GeV$ and $|\eta| < 2.4$, which reduces the background from $\ttbar$ events.

\begin{figure}[htb]
    \begin{center}
\includegraphics[width=0.45\textwidth]{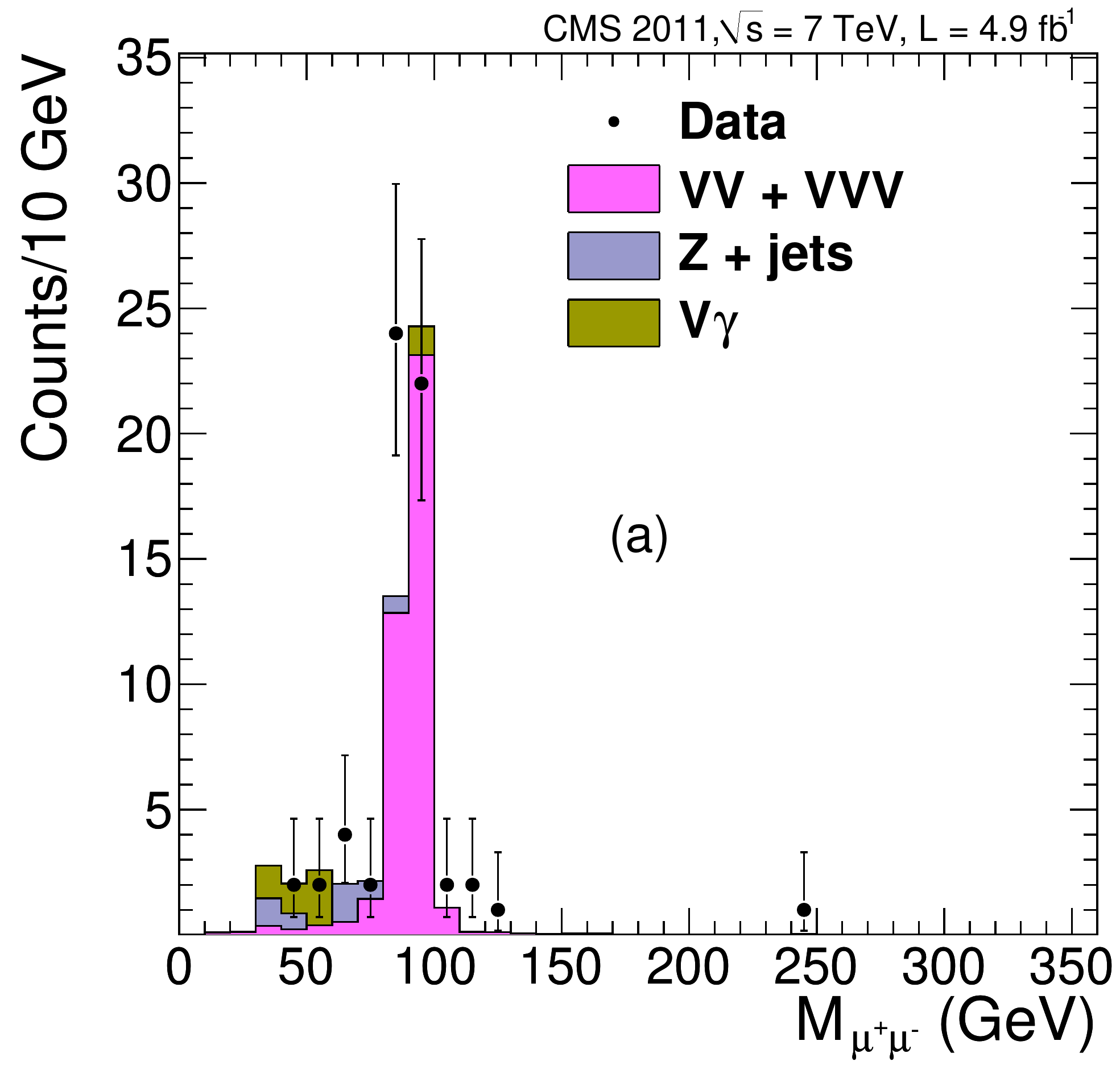}
\includegraphics[width=0.45\textwidth]{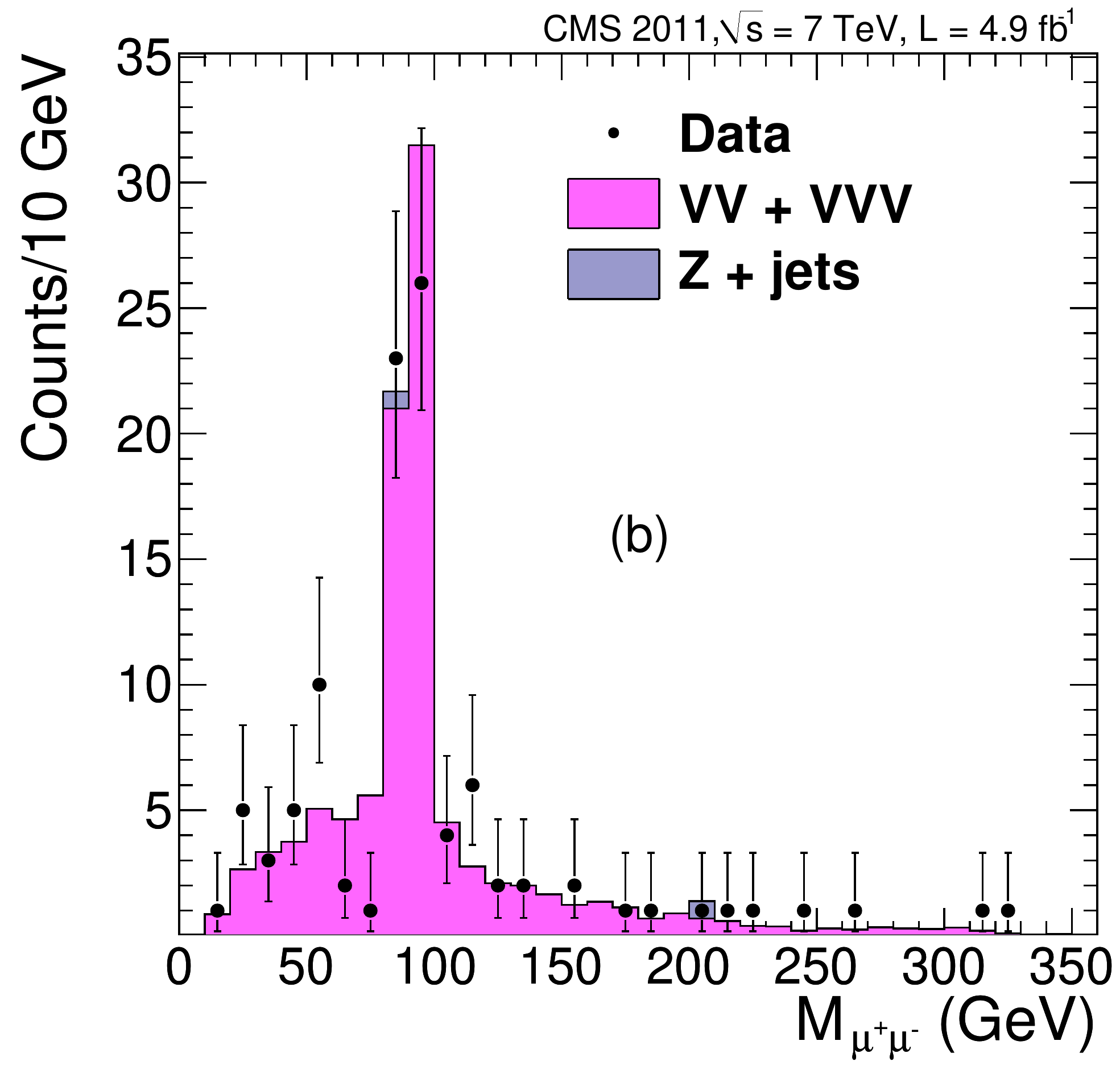}
        \caption{Distributions of the $\Pgmm\Pgmp$ invariant mass for (a)  $\Pgmm \Pep \Pgmp $ and  (b)
$\Pgmm \Pgmp \Pgmp$ events in data (black points), before applying any requirement on the $\Pgmm \Pgmp $ mass to reject Z bosons, compared to the sum of all major SM background contributions.
}
        \label{f:Mmumu}
    \end{center}
\end{figure}

The selected events
are classified into six
categories that depend on lepton flavour and electric charge:
$\Pgmm \Pep \Pep $, $\Pgmm \Pep \Pgmp $,
$\Pgmm \Pgmp \Pgmp$, $\Pem\Pgmp\Pgmp$, $\Pem\Pep\Pgmp$,
and $\Pem\Pep\Pep$.
Except for the first and fourth categories, such configurations can also result from  $\PWp$Z events.
Figure~\ref{f:Mmumu} shows the distributions of the $\Pgmm\Pgmp$ invariant mass for $\Pgmm \Pep \Pgmp $ and
$\Pgmm \Pgmp \Pgmp$ events in data, before applying any requirement on the $\Pgmm \Pgmp $ mass, compared to the sum of SM background contributions.
A peak in the $\Pgmp \Pgmm$ effective mass close to that of the Z boson is evident in both
simulated events and in data. To reduce the background from  $\PWp$Z events,  a Z veto
is added to the selection requirements for the corresponding  categories as follows.
Events with at least one  $\ell^+\ell^-$  mass combination
in the range $82<m_{\ell^+\ell^-}<102$\GeV are rejected.
To reject lepton pairs from decays of heavy-flavour quarks, events with
$m_{\ell^+\ell^-} < 12$\GeV are also discarded.

Other sources of background in final states with three leptons
arise from conversions of
photons into additional $\ell^{+} \ell^{-}$ pairs through the process
$Z \rightarrow \ell^+\ell^- \gamma \rightarrow \ell^+\ell^-\ell^{'+}\ell^{'-}$.
If one of these additional leptons carries most of the momentum of the photon,
the final state can appear as a three-lepton event.
In such cases, the invariant mass of the $\ell^+\ell^-\ell^{'}$ state peaks close
to the mass of the Z boson~\cite{sus13}.
Since the probability of a
photon  conversion to electrons is higher than to muons, an additional Z veto of
$82<m_{\ell^+\ell^-\Pep}<102$\GeV
is applied to the
$\Pgmm \Pep \Pgmp $
and $\Pem\Pep\Pep$ categories to reject such events.
This is discussed further in the next section.

\section{Background estimation \label{s:background}}

Three types of  SM processes can produce a three-lepton final
state:
(i) events containing three or more prompt leptons from production and leptonic decays of two or three EW bosons.
This is referred to as irreducible background, since it corresponds to the same final states as the signal
from $\Sigma$ production, 
(ii) V$+\gamma$ and V$+\gamma^{*}$ events, where V represents any EW boson, with the accompanying photons converting to $\ell^{+} \ell^{-}$, and
(iii)  events with one or two prompt leptons and additional non-prompt
leptons that arise from leptonic decays of hadrons within jets, called "misidentified jets".

The irreducible background from more than two leptons is dominated
by SM WZ production, but also includes ZZ and three-boson events.
The two-boson contribution, which is reduced substantially by the Z mass veto, and the three-boson contribution, which is dominated
by  the WWW channel, are both evaluated using MC simulation.
The contribution from three-boson production is small relative to the other sources, as shown in Table~\ref{tab:bckg}.

\begin{table*}[htb]
\begin{center}
\topcaption[bckg]{Summary of the mean
number of SM background events expected in each event category, after final selections.
V represents a Z or a W bosons and
V$\gamma$ is the contribution from external photon conversions.
The column labelled "Misidentified jets" includes backgrounds  with non-prompt leptons,
the column $\gamma^*\to \Pgmp \Pgmm$ shows background expectation from internal photon conversions, where a virtual photon converts to a muon pair,
and one muon is lost.
The contribution of $\gamma^*\to \Pep \Pem$  is removed by the rejection criteria on
three-lepton masses.
Statistical uncertainties are included for the six categories, and
systematic uncertainties on normalizations are listed in the last row.
}
\label{tab:bckg}
\begin{tabular}{l|c c c c c c}
\hline
                       & VV             & VVV           & V$\gamma$ &  Misidentified jets &$\gamma^*\to \Pgmp \Pgmm$\\
\hline
$\Pgmm \Pep \Pep $  & 0.3$\pm$0.1  & 0.09$\pm$0.01 & -             & 0.4$\pm$ 0.4  & -    \\
$\Pgmm \Pep \Pgmp $       & 4.0$\pm$0.3   & 0.19$\pm$0.01 & -             & 3.1$\pm$1.2  &- \\
$\Pgmm \Pgmp \Pgmp$             & 4.9$\pm$0.3    & 0.11$\pm$0.01 & -             & 5.7$\pm$1.9  &0.7$\pm$ 0.2 \\
$\Pem\Pgmp\Pgmp$          & 0.3$\pm$0.1  & 0.09$\pm$0.01 & -             & 0.8$\pm$0.5  & -  \\
$\Pem\Pep\Pgmp$     & 4.9$\pm$0.3    &0.21$\pm$0.02  & -             & 3.0$\pm$1.2  &0.4$\pm$ 0.1 \\
$\Pem\Pep\Pep$& 2.5$\pm$0.2  & 0.06$\pm$0.01 & 1.4$\pm$1.0 & 1.1$\pm$0.6  & -  \\
\hline
\footnotesize{Normalization}     & & & & &                       \\
\footnotesize{uncertainties}  & 17\% (WZ) 7.5\% (ZZ)      & 50\%  & 13\%     &  50\%              & 50\% \\
\hline
\end{tabular}
\end{center}
\end{table*}

As mentioned in Section~\ref{s:datasets}, photon conversions in the presence of W or Z bosons can produce
isolated leptons that constitute another source of background.
External conversions of photons, namely of produced photons that interact with the material in the
detector to yield primarily $\Pep\Pem$ pairs, are evaluated from simulation (V$\gamma$ in Table~\ref{tab:bckg}).
Internal conversions,
involving the direct materialisation of virtual photons into
$\Pgmp\Pgmm$ or $\Pep\Pem$ pairs, can also provide
a similar source of background. Both external and internal conversions can become problematic
when one of the two
final-state leptons carries off most of the photon energy, and the second lepton is not detected.
The contribution of conversions to electrons is reduced by the additional
three-lepton-mass
rejection applied to the
$\Pgmm \Pep \Pgmp $ and  $\Pem\Pep\Pep$ categories as discussed above.
The contribution from internal photon conversions to muons $\gamma^*\to \Pgmp \Pgmm$ is evaluated
according to the method described in Ref.~\cite{sus13}, where the ratio of $\ell^+\ell^-\mu^{\pm}$ to $\ell^+\ell^-\gamma$
events, in which the
mass is close to that of a Z boson, defines a conversion factor $C_{\mu}$  for muons.
The background is estimated from $C_{\mu}$ and from the number of $\ell^+\ell^- \gamma$ events in data that pass all selections, except the three-lepton requirements.
An alternative evaluation is obtained from events
in an independent Z-enriched control region, by reversing the $\MET$ requirement to  $\MET<20$\GeV.
As mentioned before, events from Z decays into two muons or two electrons that contain an additional muon from internal photon conversion,
produce a peak in the three-lepton invariant mass distribution close to the Z mass. The number of events
expected in the final sample is estimated from the ratio of simulated events for Z production with
$\MET>30$\GeV to that with $\MET<20$\GeV.
This estimate agrees
with that of the previous method. The $\gamma^*\to \Pgmp \Pgmm$ background contribution  is small, as can be seen in Table~\ref{tab:bckg}.
An overall
uncertainty of $\pm$ 50$\%$ is assumed for this source of background, which is limited by the statistical precision of both estimates ($30\%$), and has an
additional contribution from  the choice of normalization criteria ($40 \%$).

The largest background, aside from the irreducible backgrounds, arises from
the Z+jets process (including the Drell--Yan contribution), in which the Z boson decays leptonically,
and a jet in the event is misidentified as a third lepton.
Processes with
non-prompt leptons from heavy-flavour decays are not
simulated with sufficient accuracy with the MC generators and we therefore
use a method based on data to estimate this contribution.
The yield of such background in data is estimated using a sample of leptons that pass less restrictive selection criteria than the ones described previously.
The lepton candidates passing all selection criteria are called "tight leptons",
while those passing all but the isolation requirements are called "loose leptons".
The probability for a non-prompt lepton to pass tight selection is called the misidentification rate,
and it is measured in samples of multijet events
where a negligible fraction of the lepton candidates is expected to be due to prompt leptons.
The  contribution
to the background is obtained from the lepton misidentification rate and the  events that pass full selection of the analysis,
based on loose lepton identification.
The misidentification rate depends on
$\pt$  and $\eta$ of the lepton.
However, only the average value is used, and an uncertainty of 50$\%$ is assigned to this background estimate.
Several cross checks of the method used to evaluate this background contribution have
been performed using data and simulation. They show agreement between the
number of observed leptons and the number of leptons predicted on the basis of the lepton misidentification
rate.

Events from $\ttbar$ production with two leptonic W decays and an additional coincident lepton, are reduced through the PF isolation requirements for leptons and  by the selection on $H_\mathrm{T}$. Simulations show that the remaining $\ttbar$ background is negligible, and its contribution
is included in the estimate of non-prompt leptons.

SM background contributions expected in each of the six analyzed event categories are summarized in
Table~\ref{tab:bckg}.

\section{Systematic uncertainties \label{s:syst}}
Systematic uncertainties can be divided in two categories:
those related to the extraction of the signal and those relevant to the sources of background.
The first group includes efficiencies of trigger selections, particle reconstruction, and lepton identification.
In the kinematic region defined by the analysis, the trigger efficiency for the signal
is very high because
it is based on a combination of three separate two-lepton triggers,
each of which is found to be $92\%$ to $100\%$ efficient, and the estimated overall efficiency
is $(99 \pm 1)\%$.

Uncertainties on lepton selection efficiencies are determined using a ``tag-and-probe" method~\cite{CMSWandZ},
both in data
and through MC simulations, and the differences between these are taken
as systematic uncertainties on the efficiencies.
Additional contributions include uncertainties on the energy scales and on resolutions for leptons and for $\MET $, as well as
uncertainties in the modeling of pileup, all of which are obtained from a full \GEANTfour simulation.
As mentioned in Section~\ref{s:sim}, \GEANTfour simulation of the signal is restricted to a limited number
of $\MSig$ masses.
In fact, the largest available value for this simulation is $\MSig = 140\GeV$.
The efficiencies are therefore extrapolated to higher mass points using fast detector simulation.
The difference between
the efficiencies evaluated with the full  and fast simulation at 140\GeV
is taken as an additional contribution to the overall uncertainty.
The largest difference is for the channel  with three muons.
Statistical uncertainties of the extrapolation are also taken into account.
The uncertainties attributed to the expected signal efficiencies are
summarized in Table~\ref{tab:signalSystEff} for $\MSig=180$\GeV,
and are expected not to differ significantly for higher mass points \cite{carflo_}.

\begin{table*}[htb]
\begin{center}
\topcaption[Systematicncertainties on signal]{Uncertainties on
     signal efficiency for each event category for $\MSig =$ 180\GeV.
Total systematic and total systematic + statistical (fourth and sixth columns) are calculated in quadrature.
}
\label{tab:signalSystEff}
\begin{tabular}{l|ccc|c|cc}
\hline
 & \multicolumn{6}{c}{Source of uncertainty} \\
 &\footnotesize{Trigger}&\footnotesize{Signal efficiency}&\footnotesize{(Fullsim/Fastsim)}&\footnotesize{Total}&\footnotesize{(Fullsim/Fastsim)}&\footnotesize{Total}\\
 &          &\footnotesize{(Full simulation)}&\footnotesize{systematic}&\footnotesize{systematic}&\footnotesize{statistical}&\footnotesize{syst.+stat.}\\
\hline
$\Pgmm \Pep \Pep $   & 1.0\% & 6.3\% &  2.9\% &  7.0\% & 3.0\% &  7.6\% \\
$\Pgmm \Pep \Pgmp $        & 1.0\% & 4.5\% &  6.8\% &  8.2\% & 2.3\% &  8.5\% \\
$\Pgmm \Pgmp \Pgmp$              & 1.0\% & 3.9\% & 11.1\% & 11.8\% & 3.3\% & 12.2\% \\
$\Pem\Pgmp\Pgmp$           & 1.0\% & 4.5\% &  8.5\% &  9.7\% & 2.9\% & 10.1\% \\
$\Pem\Pep\Pgmp$      & 1.0\% & 6.3\% &  4.1\% &  7.6\% & 2.4\% &  7.9\% \\
$\Pem\Pep\Pep$ & 1.0\% & 7.6\% &  2.8\% &  8.0\% & 4.2\% &  9.1\% \\
\hline
\end{tabular}
\end{center}
\end{table*}

As mentioned above, the uncertainties on backgrounds are estimated using MC simulations or control samples in data.
For the dominant irreducible background of WZ production, we apply a 17\% uncertainty on the measured cross section~\cite{wzATLAS}.
Uncertainties of 7.5\% for ZZ~\cite{wztheory}, and 13\% for V$\gamma$~\cite{VgCMS} cross sections are also taken into account.
For very small backgrounds, such as WWW, we assume a normalization uncertainty of 50\%.

Uncertainties on background estimates from methods based on data were discussed
in Section~\ref{s:background}, and those statistical and systematic uncertainties  are summarized in Table~\ref{tab:bckg}.

The overall uncertainty on integrated luminosity is
2.2$\%$~\cite{LUMIPAS}.
For backgrounds determined from simulation,
the systematic uncertainties on  efficiency and luminosity
are common to all
signals.

\section{Results \label{s:results}}

Table~\ref{tab:AllYield} presents the results of our search for the fermionic $\Sigma$ triplet states in terms of the expected number of
signal events, the expected number of events from SM background,
and the number of observed events in each of the analyzed event categories.
Each of the three possibilities for mixing (FDS, $\mu$S, eS) described in Section~\ref{s:intro} is considered in the analysis.

\begin{table*}[htbp]
\centering
\topcaption[Final yield for signal, backgrounds and data]{Summary  of the expected
mean
 number of events for signal
as a function of $\MSig$, for the expected SM background,
and the observed number of events in data, after implementing all analysis selections.
Each of the three possibilities for mixing (FDS, $\mu$S, eS) described in Section~\ref{s:intro} are considered separately in the analysis.
}
\label{tab:AllYield}
\begin{tabular}{l|ccccc|cc|cc|c|c}
\hline
& \multicolumn{9}{c|}{Expected signal for $\MSig$\unit{(\GeVns)}} & Expected & Observed\\
& \multicolumn{5}{c|}{  FDS }& \multicolumn{2}{c|}{$\mu\text{S}$} & \multicolumn{2}{c|}{eS} & background  & in data \\
Category      & 120 & 130 & 140 & 180 & 200 & 180 & 200 & 180 & 200 & & \\
\hline
$\Pgmm \Pep \Pep $  &
  7.9 & 6.0 & 4.5 & 1.7 & 1.1 & 1.6 & 1.0 & 3.6 & 2.4 &
0.8$\pm$0.4 & 2 \\
$\Pgmm \Pep \Pgmp $       &
 12.3 &9.0 & 7.0& 3.0 & 2.0  & 6.0 & 4.0 & 1.4 & 0.92 &
7.3$\pm$2.1 & 9\\
$\Pgmm \Pgmp \Pgmp $            &
  7.8 &5.2 &3.6 &1.4 & 0.93 & 6.1& 4.0 & - & - &
11.5$\pm$3.6 & 7 \\
$\Pem\Pgmp\Pgmp$          &
  8.3 &6.2 &4.8 &1.8 & 1.2  & 3.7&  2.5 & 1.6 & 1.0 &
1.1$\pm$0.7 & 0 \\
$\Pem\Pep\Pgmp$      &
 13.2 &9.5  &6.9 &2.7 & 1.8 & 1.1&  0.75 &5.7 & 3.8 &
8.6$\pm$2.2 & 7 \\
$\Pem\Pep\Pep$ &
 3.9 &2.8 & 2.0& 1.0& 0.63 & - &  -   &4.16 & 2.8 &
5.0$\pm$1.4 & 4 \\
\hline
\end{tabular}
\end{table*}

No significant excess of events is observed relative to the SM expectations
in any of the six analysis channels. Combining all channels, we set
upper limits at the 95\% confidence level (CL) on $\sigma\times\mathcal{B}$, on the product of
the production cross section of $\Sigma^{+} \Sigma^{0}$ and its branching fraction ($\mathcal{B}$)
to the three-lepton final states, where the lepton can be an electron,
muon or $\tau$
(contributing through $\tau \to \ell \nu_{\ell} \nu_{\tau}$).
The branching fraction to three-lepton final states depends
on $\MSig$~\cite{carflo_}, and is predicted to be about 9\% for $\MSig\approx 200$\GeV, where we extrapolate signal
yields to $\MSig>180$\GeV using the results of Ref.~\cite{carflo_}.

\begin{figure}[htb]
\centering
\includegraphics[width=\cmsFigWidth]{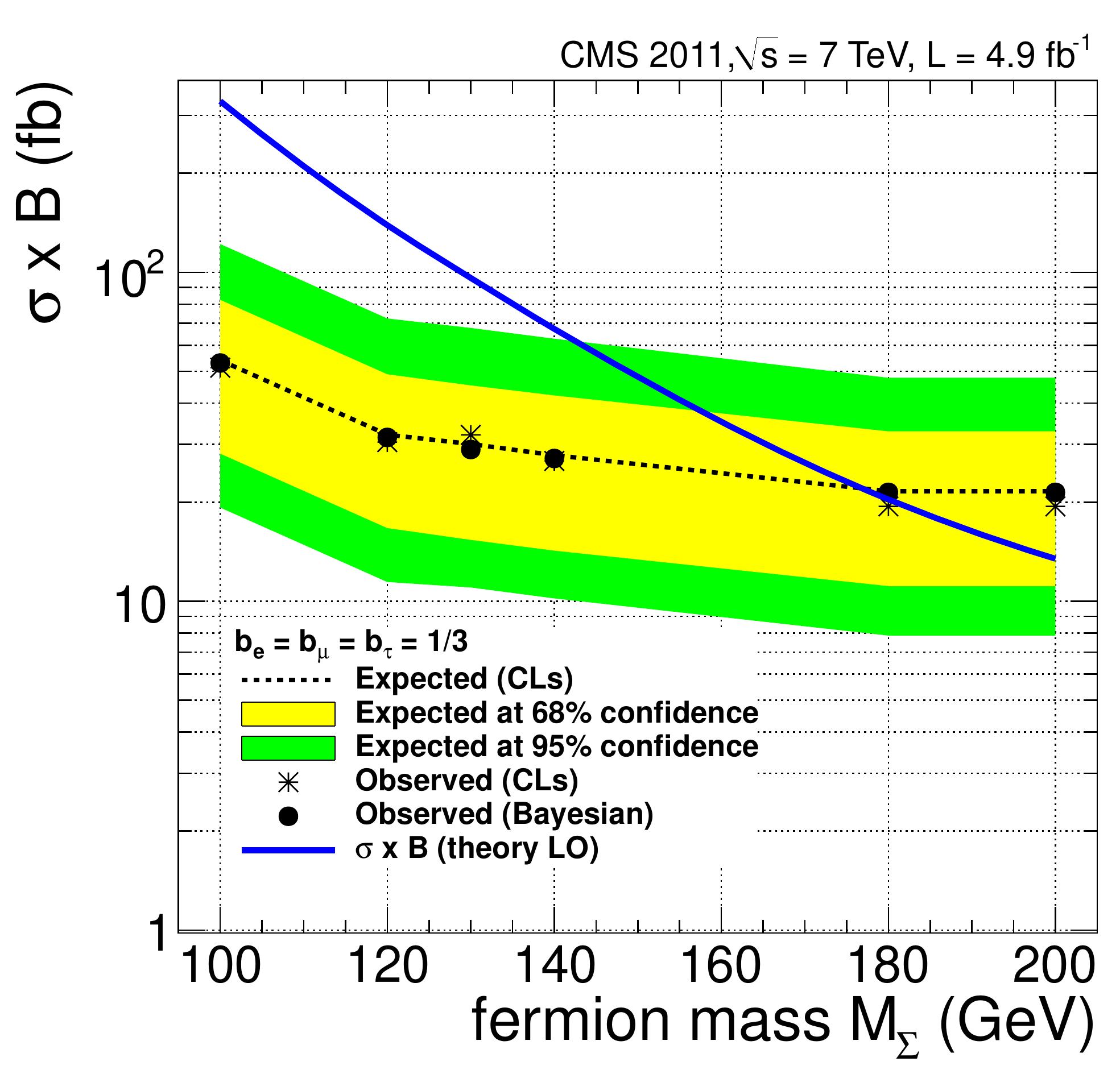}

\caption[Limit for $b_{\mu}=b_{e}=b_{\tau}=1/3$]
{The expected (dashed line) and observed (asterisks and black points) exclusion limits at 95$\%$ confidence level
on $\sigma \mathcal{B}$
as a function of the fermion mass $\MSig$, assuming
$b_{\Pe}=b_{\mu}=b_{\tau}=1/3$ (FDS) for the signal.
The solid (blue) curve represents the predictions of the LO type III seesaw models.
The light (yellow) and dark (green) shaded areas represent, respectively, the 1 standard deviation (68$\%$ CL)
and 2 standard deviations (95$\%$ CL) limits on the expected results obtained from MC pseudo-experiments,
which reflect the combined statistical and systematic uncertainties of the SM contributions.
The asterisks and the black points show, respectively, the observed limits
computed
following a frequentist method based on the CLs criterion and a Bayesian approach.
}
\label{fig:limit_Vdem}
\end{figure}

\begin{figure}[htbp]
\centering
\includegraphics[width=\cmsFigWidth]{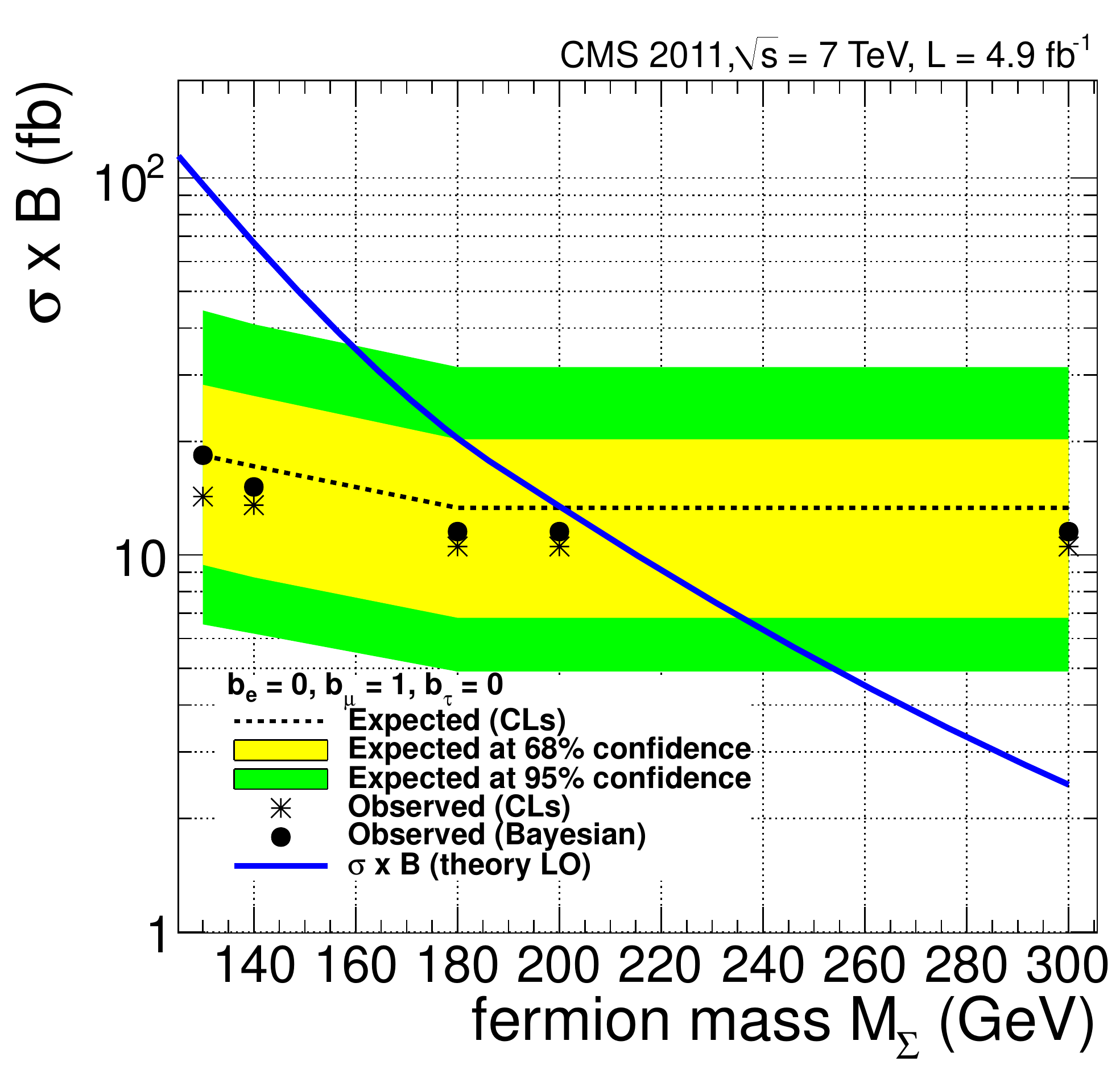}

\caption[Limit for $V_{\mu}=0.063$, and $V_{\Pe}=V_{\tau}=0$]
{The expected (dashed line) and observed (asterisks and black points) exclusion limits at 95$\%$ confidence level
on $\sigma\mathcal{B}$
as a function of the fermion mass $\MSig$, assuming $b_{\Pe}=0, b_{\mu}=1, b_{\tau}=0$ ($\mu$S) for the signal .
The solid (blue) curve represents the predictions of the LO type III seesaw models.
The light (yellow) and dark (green) shaded areas represent, respectively, the 1 standard deviation (68$\%$ CL)
and 2 standard deviations (95$\%$ CL) limits on the expected results obtained from MC pseudo-experiments,
which reflect the combined statistical and systematic uncertainties of the SM contributions.
The asterisks and the black points show, respectively, the observed limits
computed
following a frequentist method based on the CLs criterion and a Bayesian approach.
}
\label{fig:limit_Vmu}
\end{figure}

\begin{figure}[htbp]
\centering
\includegraphics[width=\cmsFigWidth]{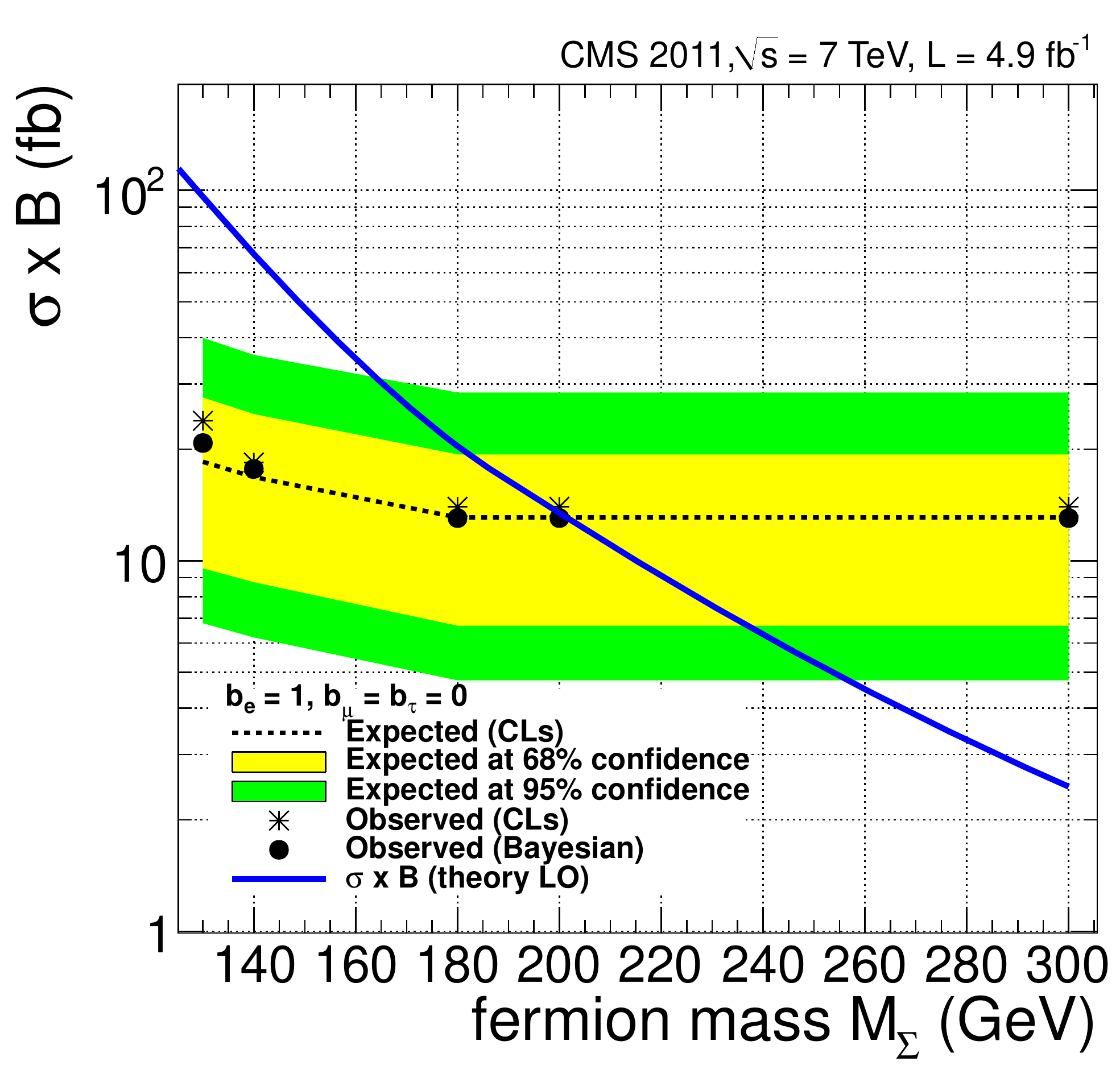}

\caption[Limit for$V_{\Pe}=0.05$ and $V_{\mu}=V_{\tau}=0$]
{The expected (dashed line) and observed (black points) exclusion limits at 95$\%$ confidence level
on $\sigma \mathcal{B}$
as a function of the fermion mass $\MSig$, assuming $b_{\Pe}= 1, b_{\mu}=0, b_{\tau}=0$ (eS) for the signal.
The solid (blue) curve represents the predictions of the LO type III seesaw models.
The light (yellow) and dark (green) shaded areas represent, respectively, the 1 standard deviation (68$\%$ CL)
and 2 standard deviations (95$\%$ CL) limits on the expected results obtained from MC pseudo-experiments,
which reflect the combined statistical and systematic uncertainties of the SM contributions.
The asterisks and the black points show, respectively, the observed limits
computed
following a frequentist method based on the CLs criterion and a Bayesian approach.
}
\label{fig:limit_Ve}
\end{figure}

The upper limits on $\sigma\mathcal{B}$ as a function of fermion mass $\MSig$,
combining for all channels by multiplying the corresponding likelihood functions,
are shown in
Fig.~\ref{fig:limit_Vdem}, ~\ref{fig:limit_Vmu}, and ~\ref{fig:limit_Ve}, for FDS, $\mu$S, and eS possibilities, respectively.
The dashed lines correspond to the expected limits obtained from MC pseudo-experiments,
and are based on the CLs criterion~\cite{Tommaso_1,Tommaso_2}.
The observed limits on data are computed following both a Bayesian approach~\cite[Ch.~33]{PDG_}, and a frequentist method also based on the CLs criterion.
In the former, the assumed prior
is a constant.
In both calculations, the  uncertainties on efficiencies for detecting signal, the uncertainty on integrated luminosity and on the
expected SM background,
are treated as uninteresting ``nuisance"  parameters with Gaussian or log-normal densities.
Upper limits are computed at 95$\%$ CL using the \textsc{RooStats} software~\cite{RooStats}, and
and the package developed to combine results from searches for the Higgs boson~\cite{CLs_}.
The two results are similar, as shown in Figs.~\ref{fig:limit_Vdem}, ~\ref{fig:limit_Vmu}, and ~\ref{fig:limit_Ve}.
The results are stable relative to variations of ${\pm}20 \%$ on the systematic uncertainties.
Finally, we extract lower limits on $\MSig$ using the theoretical dependence of the cross section on $\MSig$, as represented by the solid blue lines of Fig.~\ref{fig:limit_Vdem}, \ref{fig:limit_Vmu}, and \ref{fig:limit_Ve}, for the three possibilities for the type III seesaw model for signal.
The expected and observed 95$\%$ CL limits obtained with the Bayesian method
are given in Table~{\ref{tab:Limits}}.

\begin{table}[htbp]
\begin{center}
\topcaption[]{The expected and observed limits on $\MSig$ and on $\sigma \mathcal{B}$ at the given mass are obtained using the Bayesian method, specified at a $95\%$ confidence level, for the three assumed sets of branching fractions $b_{\alpha}$ defined in Eq.(\ref{eq:br}).}
\label{tab:Limits}
\begin{tabular}{l|cccc}
\hline
Scenario &  \multicolumn{2}{|c}{95$\%$ CL: $\sigma \mathcal{B}$ (fb)}
                & \multicolumn{2}{c}{95$\%$ CL: $\MSig\unit{(\GeVns)}$ } \\
                & Exp. & Obs. & Exp. & Obs. \\
\hline
FDS    & 22 & 20 & 177 & 179 \\
$\mu$S & 13 & 11 & 201 & 211 \\
eS     & 13 & 13 & 202 & 204 \\
\hline
\end{tabular}
\end{center}
\end{table}

The reported limits are valid only for short $\Sigma$ lifetimes, which hold for values of the matrix elements $V_{\alpha}$ greater than ${\approx}10^{-6}$.
For smaller values, the analysis requires a different approach, since the leptons can originate from displaced vertices in an environment that, as indicated previously, is not considered in this analysis.

\section{Summary \label{s:conclusions}}
A search has been presented for fermionic triplet states expected in type III seesaw models.
The search was performed in events with three isolated leptons (muons or electrons), whose charges sum to $+1$, and contain jets and an imbalance in transverse momentum.
The data are from proton-proton collisions at $\sqrt{s}=7$\TeV, recorded during 2011 by the CMS experiment at the CERN LHC, and correspond to an integrated
luminosity of 4.9\fbinv.

No evidence for  pair production of  $\Sigma^{+} \Sigma^{0}$ states has been found,
and 95$\%$ confidence upper limits are set on the product of the production cross section of $\Sigma^{+} \Sigma^{0}$
and its branching fraction to the examined three-lepton final states.
Comparing the results with predictions from type III seesaw models,
lower bounds are established at 95$\%$ confidence on the mass of the $\Sigma$ states.
Limits are reported for three choices of mixing possibilities between the $\Sigma$ states and the three lepton generations.
Depending on the considered scenarios, lower limits are obtained on the mass of the heavy partner of the neutrino that range from 180 to 210\GeV.
The results are valid only if at least one of the mixing matrix elements is larger than $\approx$ $10^{-6}$.
These are the  first limits on the production of type III seesaw fermionic
triplet  states reported by an experiment at the LHC.

\section*{Acknowledgment \label{s:ack}}

We congratulate our colleagues in the CERN accelerator departments for the excellent performance of the LHC machine. We thank the technical and administrative staffs at CERN and other CMS institutes, and acknowledge support from BMWF and FWF (Austria); FNRS and FWO (Belgium); CNPq, CAPES, FAPERJ, and FAPESP (Brazil); MEYS (Bulgaria); CERN; CAS, MoST, and NSFC (China); COLCIENCIAS (Colombia); MSES (Croatia); RPF (Cyprus); MoER, SF0690030s09 and ERDF (Estonia); Academy of Finland, MEC, and HIP (Finland); CEA and CNRS/IN2P3 (France); BMBF, DFG, and HGF (Germany); GSRT (Greece); OTKA and NKTH (Hungary); DAE and DST (India); IPM (Iran); SFI (Ireland); INFN (Italy); NRF and WCU (Republic of Korea); LAS (Lithuania); CINVESTAV, CONACYT, SEP, and UASLP-FAI (Mexico); MSI (New Zealand); PAEC (Pakistan); MSHE and NSC (Poland); FCT (Portugal); JINR (Armenia, Belarus, Georgia, Ukraine, Uzbekistan); MON, RosAtom, RAS and RFBR (Russia); MSTD (Serbia); SEIDI and CPAN (Spain); Swiss Funding Agencies (Switzerland); NSC (Taipei); ThEP, IPST and NECTEC (Thailand); TUBITAK and TAEK (Turkey); NASU (Ukraine); STFC (United Kingdom); DOE and NSF (USA).
Individuals have received support from the Marie-Curie programme and the European Research Council (European Union); the Leventis Foundation; the A.P.Sloan Foundation; the Alexander von Humboldt Foundation; the Belgian Federal Science Policy Office; the Fonds pour la Formation \`a la Recherche dans l'Industrie et dans l'Agriculture (FRIA-Belgium); the Agentschap voor Innovatie door Wetenschap en Technologie (IWT-Belgium); the Ministry of Education, Youth and Sports (MEYS) of Czech Republic; the Council of Science and Industrial Research, India; the Compagnia di San Paolo (Torino); and the HOMING PLUS programme of Foundation for Polish Science, cofinanced from European Union, Regional Development Fund.

\bibliography{auto_generated}   

\newpage

\cleardoublepage \appendix\section{The CMS Collaboration \label{app:collab}}\begin{sloppypar}\hyphenpenalty=5000\widowpenalty=500\clubpenalty=5000\textbf{Yerevan Physics Institute,  Yerevan,  Armenia}\\*[0pt]
S.~Chatrchyan, V.~Khachatryan, A.M.~Sirunyan, A.~Tumasyan
\vskip\cmsinstskip
\textbf{Institut f\"{u}r Hochenergiephysik der OeAW,  Wien,  Austria}\\*[0pt]
W.~Adam, E.~Aguilo, T.~Bergauer, M.~Dragicevic, J.~Er\"{o}, C.~Fabjan\cmsAuthorMark{1}, M.~Friedl, R.~Fr\"{u}hwirth\cmsAuthorMark{1}, V.M.~Ghete, J.~Hammer, N.~H\"{o}rmann, J.~Hrubec, M.~Jeitler\cmsAuthorMark{1}, W.~Kiesenhofer, V.~Kn\"{u}nz, M.~Krammer\cmsAuthorMark{1}, I.~Kr\"{a}tschmer, D.~Liko, I.~Mikulec, M.~Pernicka$^{\textrm{\dag}}$, B.~Rahbaran, C.~Rohringer, H.~Rohringer, R.~Sch\"{o}fbeck, J.~Strauss, A.~Taurok, W.~Waltenberger, G.~Walzel, E.~Widl, C.-E.~Wulz\cmsAuthorMark{1}
% ...
\vskip\cmsinstskip
\dag:~Deceased\\
1:~~Also at Vienna University of Technology, Vienna, Austria\\
% ...

\end{sloppypar}
\end{document}